\documentclass{aa}  

\usepackage{graphicx}
\usepackage{txfonts}
\usepackage[colorlinks=true, citecolor=blue, linkcolor=blue]{hyperref}
\usepackage{xcolor}
\usepackage[normalem]{ulem}

\usepackage{natbib}
\bibpunct{(}{)}{;}{a}{}{,} 

\begin{document}

   \title{Horizontal motions in sunspot penumbrae}

   \subtitle{}

   \author{Michal Sobotka
          \inst{1}\fnmsep\thanks{michal.sobotka@asu.cas.cz}
          \and
          Klaus G. Puschmann \inst{2}
          }

   \institute{Astronomical Institute of the Czech Academy of Sciences (v.v.i.),
              Fri\v{c}ova 298, 25165 Ond\v{r}ejov, Czech Republic
   \and
    Berlin, Germany
 }

  \date{Received 17 March 2022; accepted 13 April 2022}


\abstract
  {A model of penumbral filaments represented by magnetoconvective cells was derived recently from spectropolarimetric observations. This model resolves many of the inconsistencies found in the relations between intensity, magnetic, and velocity patterns in sunspot penumbrae.}
   {High-resolution observations of horizontal motions in the penumbra are needed to complement the concept of penumbrae obtained from spectropolarimetry. Time series of intensity images of a large sunspot in AR 10634 acquired with the Swedish Solar Telescope in the G band and red continuum are analysed. The two simultaneous time series last six hours and five minutes.}
   {Horizontal motions of penumbral grains (PGs), structures in dark bodies of filaments, the outer penumbral border, and G-band bright points are measured in time slices that cover the whole width of the penumbra and the neighbouring granulation. The spatial and temporal resolutions are 90~km and 20.1~s, respectively.}
   {In the inner penumbra, PGs move towards the umbra (inwards) with a mean speed of $-0.7$~km\,s$^{-1}$. The direction of motion changes from inwards to outwards at approximately 60\% of the penumbral width, and the mean speed increases gradually in the outer penumbra, approaching 0.5~km\,s$^{-1}$. This speed is also typical of an expansion of the penumbra--granulation border during periods that typically last one hour and are followed by a fast contraction. The majority of the G-band bright points moves away from the sunspot, with a typical speed of 0.6~km\,s$^{-1}$. High outward speeds, 3.6~km\,s$^{-1}$ on average, are observed in dark bodies of penumbral filaments.}
   {According to the model of penumbral filaments, it is suggested that the speeds detected in the dark bodies of filaments are associated with the Evershed flow and that the opposite directions of PG motions in the inner and outer penumbrae may be explained by the interaction of rising plasma in filament heads with a surrounding, differently inclined magnetic field.}


  \titlerunning{Horizontal motions in sunspot penumbrae}

\maketitle
%

\section{Introduction}
\label{sec:intro}

A sunspot penumbra is formed in the presence of a strong and inclined magnetic field. To explain its observed brightness, some kind of convection must take place within it. This implies a system of vertical and horizontal flows. Upflows are observed predominately in the inner penumbra, while downflows are concentrated at the outer penumbral boundary \citep[e.g.][]{Franz2009}. This results in a strong horizontal radial outflow, as discovered by \citet{Evershed1909}. 
The magnetic field in the penumbra can be characterised by two interlacing systems of field lines: one more vertical and one more horizontal \citep{SolankiMont1993}. Radially elongated parts of the penumbra with a more vertical and stronger field, called spines \citep{Lites1993}, alternate with a nearly horizontal field in interspines. The more vertical field wraps around the more inclined one, and the spines seem to be a continuation of the umbral magnetic field \citep{Tiwari2015}. 
\citet{Schlich1998ApJ, Schlich1998AA} proposed a dynamical model of a flux tube in the interspine with a hot plasma upflow at a place observed as a bright head of a penumbral filament, oriented towards the umbra: the penumbral grain \citep[PG;][]{Muller1973}. In the photosphere, the hot plasma inside the tube is cooled by radiation as it flows horizontally outwards. As long as the flowing plasma is hotter than its surroundings, it constitutes a bright radial filament; however, at a longer distance the filament becomes gradually dark thanks to the radiative cooling. The tube constitutes a channel that carries the Evershed flow.

The classical picture of sunspot penumbrae consisting of bright and dark fibrils suffered from a number of problems and disagreements, as described, for example, in the extended reviews by \citet{Solanki2003}, \citet{BorreroIchimoto2011}, and \citet{Tiwari2017}. Particularly, the relations of the intensity structures to magnetic field and Evershed flow channels were found to be weak and unclear. 
Recently, \citet{Tiwari2013} presented a model of a penumbral filament derived from inversions of Hinode spectropolarimetric observations. This filament represents a magnetoconvective cell, where hot plasma from sub-photospheric layers enters into the filament's head (the PG), which is the hottest and brightest part of the filament, with a magnetic field weaker than that in its surroundings (spines) and moderately inclined. The mass upflow of about 5~km\,s$^{-1}$ in the head then becomes horizontal in the filament's body, parallel to the $\sim$1~kG horizontal magnetic field, and leaks laterally, producing narrow downflow regions on the sides of the filament. The radial flow along the filament's axis generates the Evershed effect. The filament's body becomes gradually darker than its surroundings due to radiative cooling. The flow turns down at the end of the filament, producing a strong downflow of 7~km\,s$^{-1}$ and a local heating in deep photospheric layers. Tiwari's picture of a sunspot penumbra consisting of convective filaments and spines resolves most of the past issues. However, the volution of filaments and related horizontal motions are missing from this picture due to the limited cadence of Hinode spectropolarimetric data.

Measurements of horizontal (or proper) motions, using high-cadence data, can complement the concept of penumbrae obtained from the spectropolarimetric observations discussed above. Different methods can be applied to time series of intensity images or Dopplergrams to measure speeds of horizontal motions: local correlation tracking \citep[LCT,][]{NovSim1988}, feature tracking \citep[e.g.][]{SBSIII1999}, and time slices.  It is necessary to note that the proper motions of intensity or velocity structures may not directly correspond to real plasma flows.

The LCT method suffers from inherent space and time averaging, which reduce the observed velocities, and from its inability to distinguish between motions of bright and dark features. The LCT results of \citet{WangZ1992}, \citet{Denker1998}, and recently \citet{Vafa2015} have revealed that PGs and dark features in the inner penumbra move towards the umbra (inwards) at a typical speed of 0.5~km\,s$^{-1}$ with a maximum of 0.7~km\,s$^{-1}$, and the elements in the outer penumbra move outwards at approximately the same speed with a maximum of 2~km\,s$^{-1}$. The inward motion of PGs, reported already by \citet{Muller1973}, was explained in terms of the dynamical flux-tube model \citep{Schlich1998AA}. The rise of the flux tube causes the crossing point of the tube with the visible surface, observed as a PG, to move radially inwards. It is clear that this motion does not correspond to the upward- and outward-oriented plasma flow in the tube, so the observed inward motions of PGs are apparent.

The feature-tracking method makes it possible to distinguish between different types of features. It relies on a correct image segmentation that isolates the objects under study. \citet{SBSIII1999} and \citet{SSIV2001} tracked bright features identified as PGs in high-resolution ground-based observations. They found inward- as well as outward-moving PGs with mean speeds of 0.4~km\,s$^{-1}$ and 0.5--0.8~km\,s$^{-1}$, respectively. The inward-moving PGs dominated the inner penumbra, while the majority of outward-moving PGs appeared in the outer penumbra. These findings were later confirmed by \citet{Zhang2013}, who applied their feature-tracking routines to Hinode observations from space. They reported mean speeds 0.7~km\,s$^{-1}$ for the inward-moving PGs and 0.65~km\,s$^{-1}$ for the outward-moving ones. In principle, the feature-tracking method could also be utilised to measure motions of dark features, that is, the bodies of penumbral filaments. However, unlike PGs, dark features in the penumbra have diffuse, largely elongated shapes, and they do not show clearly defined intensity minima; these properties make the tracking unreliable.

Time slices are an alternative to the two above-mentioned methods. They can be easily made from a time series of images by fixing a certain spatial track in the image plane and mapping the time evolution of values along this track. Time slices are  2D images, where the horizontal usually shows positions along the spatial track and the vertical corresponds to elapsed time. Although the time slices were widely used to depict oscillatory and wave phenomena, they were seldom applied to the evolution of penumbral structure. \citet{Shineetal1994} used this method to study motions of PGs and places with an increased Doppler signal of Evershed flow (`Evershed clouds') along several penumbral filaments in series of continuum images and Dopplergrams. They detected the inward motion of PGs (0.5~km\,s$^{-1}$) in the inner penumbra, the fast outward motion ($\sim$4~km\,s$^{-1}$) of places with increased Doppler signal in the outer penumbra, and advection flows ($\sim$1~km\,s$^{-1}$) away from the penumbra in the neighbouring granulation.
Proper motions of Evershed clouds were also measured by \citet{Cabreraetal2007} in a time series of Dopplergrams. They obtained outward-oriented speeds in the range 2--4~km\,s$^{-1}$.

In this work we apply the time-slice method to a six-hour-long time series of intensity images with high spatial and temporal resolution to measure proper motions of PGs, dark structures in the filaments' bodies, and G-band bright points (GBPs) in the neighbouring granulation. The length of the series also makes it possible to measure the expansion or contraction of the penumbra--granulation border. We have to note that spectropolarimetric observations were not made in our observing run, so magnetic-field parameters, temperature maps, and Dopplergrams are not used in our study.


\section{Observations and data processing} \label{sec:obs}

We observed a large sunspot with the 1-metre Swedish Solar Telescope \citep{SchaSST2003}, equipped with adaptive optics \citep{SchaAO2003}, under excellent seeing conditions on 18 June 2004. The leading sunspot in AR 10634 was growing to the maximum phase of evolution, and it was quite stable during the observing period, which took place from 07:43 to 15:30 UT. It was located close to the centre of the solar disc (N13, E12, heliocentric angle $\vartheta = 13.5\degr$, $\cos{\vartheta} = 0.97$), so observed proper motions of sunspot structures were only slightly influenced by the projection effect and were practically equal to horizontal motions. 

Broadband images of the spot were acquired simultaneously in the G band ($430.89 \pm 0.6$~nm), blue continuum ($450.8 \pm 0.5$~nm), and red continuum ($602.0 \pm 1.3$~nm). Exposure times were 11--14~ms, and the pixel size was 0\farcs 0405 $\times$ 0\farcs 0405. Image acquisition in the frame-selection mode resulted in an average time difference between two selected frames of 20.1~s, which is the time resolution of our observation. The intensities in each image were calibrated using dark and flat-field frames and then normalised to the average intensity, $I_{\rm phot}$, of a quiet granulation area near the spot to compensate for changes of transparency and exposure time. The images were corrected for scattered light in the Earth's atmosphere and telescope, and the instrumental profile of the diffraction-limited 1-metre telescope was deconvolved simultaneously with noise filtering, using Wiener filters. The noise suppression started at 0\farcs 13 for the G band, 0\farcs 11 for the blue continuum, and 0\farcs 14 for the red continuum. Regarding the wavefront correction done by the adaptive optics, these values characterise the spatial resolution in the best frames. The deconvolved frames were of very good quality, and further image reconstruction was not necessary. Then, the image rotation was compensated for and the frames were aligned and de-stretched. Five-minute oscillations and a residual jitter caused by the seeing were removed by a subsonic filter with a cutoff at 4~km\,s$^{-1}$. The data reduction has been described in detail by \citet{SobPusch2009}, who used this dataset for a study of fine structures in the umbra.

\begin{figure}[t]\centering
\includegraphics[width=0.49\textwidth]{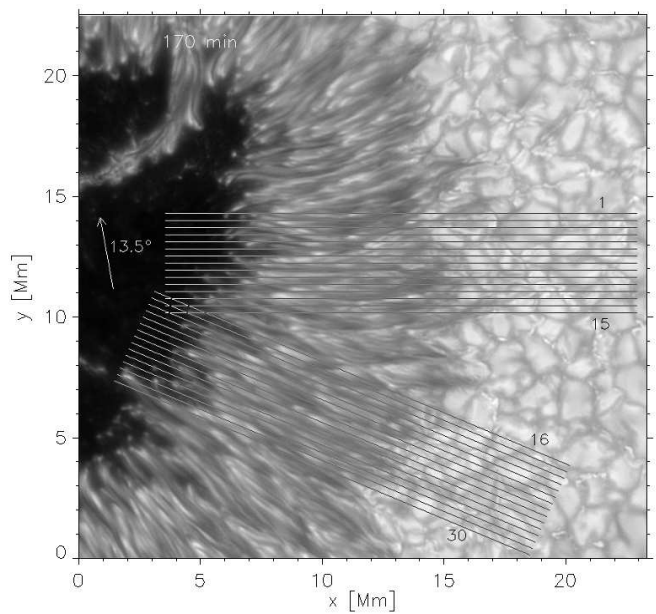}
\caption{Penumbra and neighbouring granulation of the sunspot observed in the red continuum, approximately in the middle of the series. Straight lines depict positions of time slices, identified by numbers 1--30. The distance between two neighbouring lines is 0\farcs 4. The arrow points to the disc centre. The temporal evolution of structures in the red continuum and G band is shown in two movies, available online.
\label{fig:slicepos}}
\end{figure}

We selected the series of 1090 red-continuum and 1090 \mbox{G-band} images acquired from 08:05 to 14:10~UT for the study of horizontal motions in the penumbra. The G-band images were re-aligned to the red ones, and a field of view of $32'' \times 31''$ was selected to cover a part of the umbra, the whole width of the penumbra, and the neighbouring granulation (Fig.~\ref{fig:slicepos}). Movies in both wavelength bands were created and are available online.

\begin{figure*}[t]\centering
\includegraphics[width=0.9\textwidth]{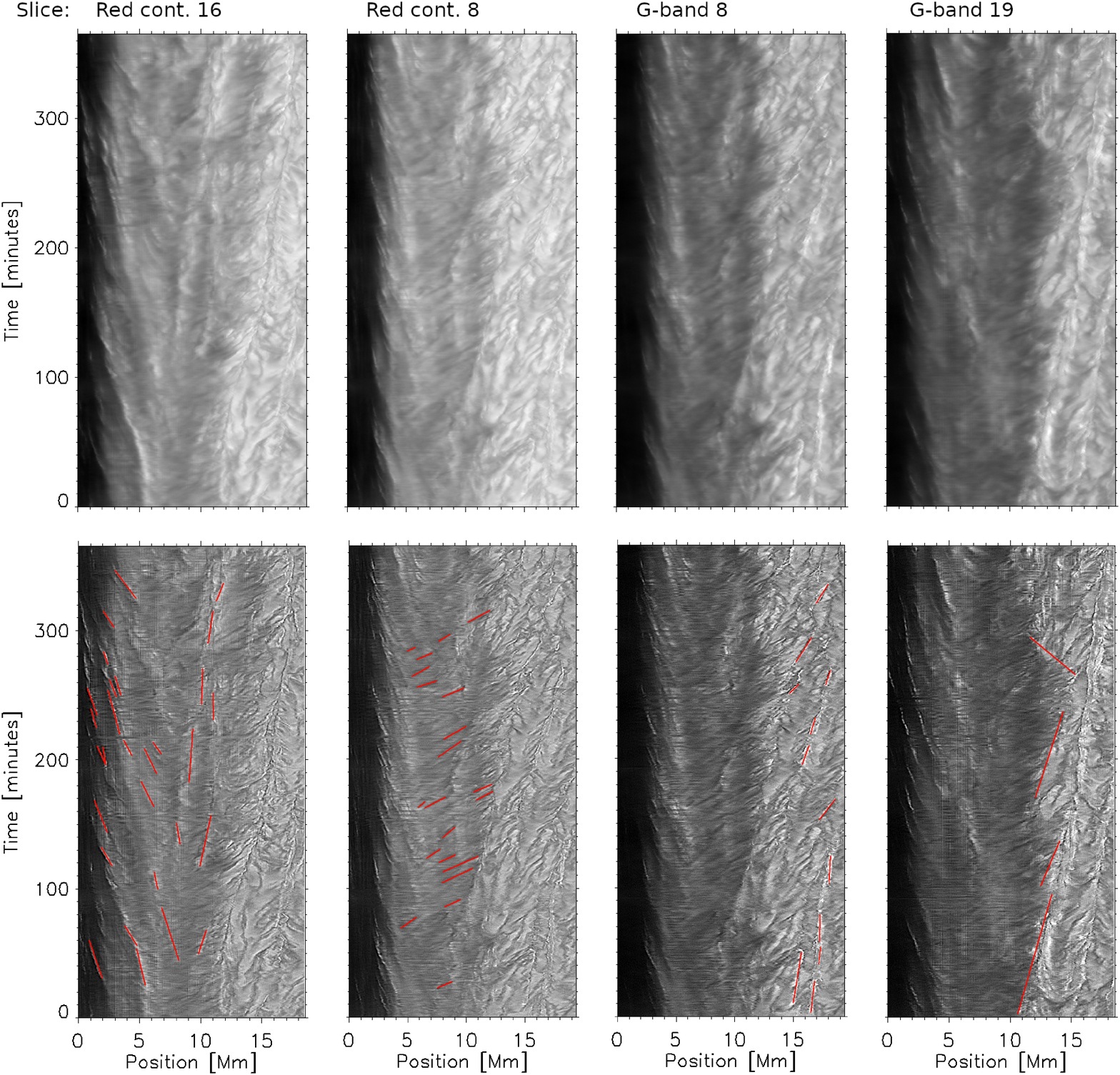}
\caption{Examples of time slices. Slices 16 and 8 (red continuum) show traces of bright PGs and dark bodies of filaments, respectively. Slices 8 and 19 (G band) show traces of GBPs and the expansion and contraction of the penumbra--granulation border, respectively.
\mbox{{\it Top row}}: Original time slices. {\it Bottom row}: Time slices enhanced by unsharp masking together with red lines drawn manually to measure speeds and durations of features.
\label{fig:sliceexamp}}
\end{figure*}

In an ideal case, a spatial track in the images that defines the time slice should follow the shape of a filament. Penumbral filaments are usually curved, and even though it would be possible to make a curved track along the filament axis, the filaments evolve but the track is fixed in time. This means that different structures can enter the spatial track throughout the time series. Moreover, it is difficult to distinguish dark bodies of filaments from spines when the information about magnetic and Doppler velocity fields is missing \citep[cf.][]{Tiwari2013}. To circumvent these problems, we defined spatial tracks as groups of densely spaced, straight parallel lines oriented approximately parallel to the filaments' directions. Two such groups, each of 15 lines, are shown in Fig.~\ref{fig:slicepos}. The distance between two neighbouring lines is 0\farcs 4, which is half of the typical filament width according to \citet{Tiwari2013}.

The group of lines covers an area where different features, for example parts of filaments, are evolving. They coincide with individual lines more or less randomly, but, in sum, nearly all of them form traces in the set of time slices of the group. These traces look like inclined bright or dark paths, the inclination of which corresponds to the speed of motion, and the vertical extent characterises the time period when a feature coincides in position with the line. This duration of a feature in a time slice is usually shorter than its real lifetime. Although in many cases an individual trace gives incomplete information on the speed and lifetime of a given feature, a statistic of numerous measurements can provide plausible results.
We created the time slices from 3D data cubes of red and G-band intensity movies. To reduce noise, the intensities at each position along each line were averaged over an area of $3 \times 3$ pixels (0\farcs 12)$^2$, which is comparable to the resolution limit. The lines begin in the umbra, cross the whole width of the penumbra, and end in the granulation (Fig.~\ref{fig:slicepos}). Their lengths are 26\farcs 7 (19.4~Mm, step 29~km) for the group of horizontal slices, 1--15, and 25\farcs 5 (18.5~Mm, step 32~km) for the group of inclined slices, 16--30.

We measured the speeds of horizontal motions and durations (lifetimes) of four types of features in the time slices. The red-continuum series was used to examine (1) PGs and (2) structures in dark bodies of penumbral filaments. The G-band series was used for measurements of (3) motions of the outer penumbral border and (4) photospheric GBPs. All the features are seen and coincide very well in both wavelength bands, but (1) and (2) are more visible in the less noisy red band and (3) and (4) have a higher contrast in the G band. 

The measurements were made visually using lines drawn over the position--time traces of the features in the time slices. We took only clearly defined traces into account and used images of the time slices enhanced by unsharp masking to draw the lines. Some examples are shown in Fig.~\ref{fig:sliceexamp}. The top row from left to right shows time slices 16 (red), 8 (red), 8 (G band), and 19 (G band), with well-marked traces of PGs, dark bodies of filaments, GBPs, and positional changes of the outer penumbral border, respectively. The corresponding enhanced images together with the lines used to measure speeds and durations are displayed in the bottom row. These examples show only one type of feature in each individual slice for better clarity, but we measured all types of features in each time slice if their well-defined traces were present. Comparing the values of spatial (0\farcs 12 $\simeq$ 90~km) and time (20.1~s) resolutions with typical travel distances and durations of features, the relative error of an individual measurement is estimated to be smaller than 10\,\%.


\section{Results} \label{sec:results}

\begin{table}\centering
\caption{Mean velocities and lifetimes.}  \label{tab:tabvel}
\begin{tabular}{lrrc}
\hline\hline
Feature  & $n$ & $v$ [km\,s$^{-1}$]  & $t$ [minutes] \\
\hline
PGs, inner penumbra       & 546 & $-0.7 \pm 0.3$ & $19 \pm 8$ \\
PGs, outer penumbra       & 126 & $-0.1 \pm 0.7$ & $20 \pm 9$ \\
Dark bodies of filaments  & 566 & $3.6 \pm 0.9$  & $7 \pm 3 $ \\
Border expansion          &  74 & $0.5 \pm 0.2$  & $60 \pm 23$\\
Border stationary         &  18 & $0.0 \pm 0.1$  & $68 \pm 32$\\
Border contraction        &  20 & $-2.2 \pm 1.3$ & $19 \pm 9$ \\
G-band bright points      & 241 & $0.8 \pm 0.7$  & $16 \pm 9$ \\
\hline
\end{tabular}
\tablefoot{Symbols: $n$ is the number of measurements, and $v$ and $t$ are the mean horizontal speed and duration of the features, respectively. A negative sign for $v$ indicates motion towards the umbra, positive away from the umbra.}
\end{table}

Mean values of horizontal speeds and durations of different features are summarised in Table~\ref{tab:tabvel}. Standard deviations characterise the scatter of individual measurements. We assigned negative values of speeds to the features moving towards the umbra (inwards) and positive to those moving away from it (outwards). The motions of PGs located in the inner half of the penumbra are directed inwards with an average speed of $-0.7$~km\,s$^{-1}$, while PGs in the outer half of the penumbra move inwards and outwards, showing a larger scatter of speeds. The mean duration time of PGs of approximately 20 minutes is the same in the inner and outer penumbrae. Dark features in the bodies of filaments always move outwards with speeds in the range 2--6~km\,s$^{-1}$, and the mean speed is 3.6~km\,s$^{-1}$. Their mean duration, 7 minutes, is substantially shorter than that of PGs. Their traces often appear repeatedly in time slices at approximately the same position, with quasi-periods of 10 and 20~minutes estimated from Fourier analysis. This effect was already found by \citet{Shineetal1994} in time slices of Dopplergrams.

It can be observed in the movies that the position of the penumbra--granulation boundary changes with time. The length of the time series of 365 minutes makes it possible to follow these long-term changes. The penumbral border tends to expand gradually, with a speed of 0.5~km\,s$^{-1}$, over one hour on average and then shrinks rapidly when a granule or several granules are formed nearby. This is demonstrated in the example shown in the rightmost column of Fig.~\ref{fig:sliceexamp}. Sometimes the border remains stationary. In Table~\ref{tab:tabvel} we present, separately, the mean values for the expansion of the penumbral border (speed $v > 0.2$~km\,s$^{-1}$),  the contraction of the penumbral border ($v < -0.2$~km\,s$^{-1}$) , and the stationary case ($-0.2 < v < 0.2$~km\,s$^{-1}$).

The tracking of GBPs provides complementary information about motions of magnetic elements in the vicinity of a sunspot. It can be seen in the G-band movie that many bright points appear close to the penumbral border, sometimes at the ends of penumbral filaments \citep{Bonetal2004}, and they follow the expansion of the border. Other GBPs appear in intergranular lanes, and their motions are influenced by the evolution of neighbouring granules, including the general outflow of granules in the sunspot moat \citep{Muller1987, Bonetal2005}. Speeds of GBPs observed in the time slices show a large scatter due to the interaction with granules, but their mean value, 0.8~km\,s$^{-1}$, is consistent with previous reports.

\begin{figure}[t]\centering
\includegraphics[width=0.5\textwidth]{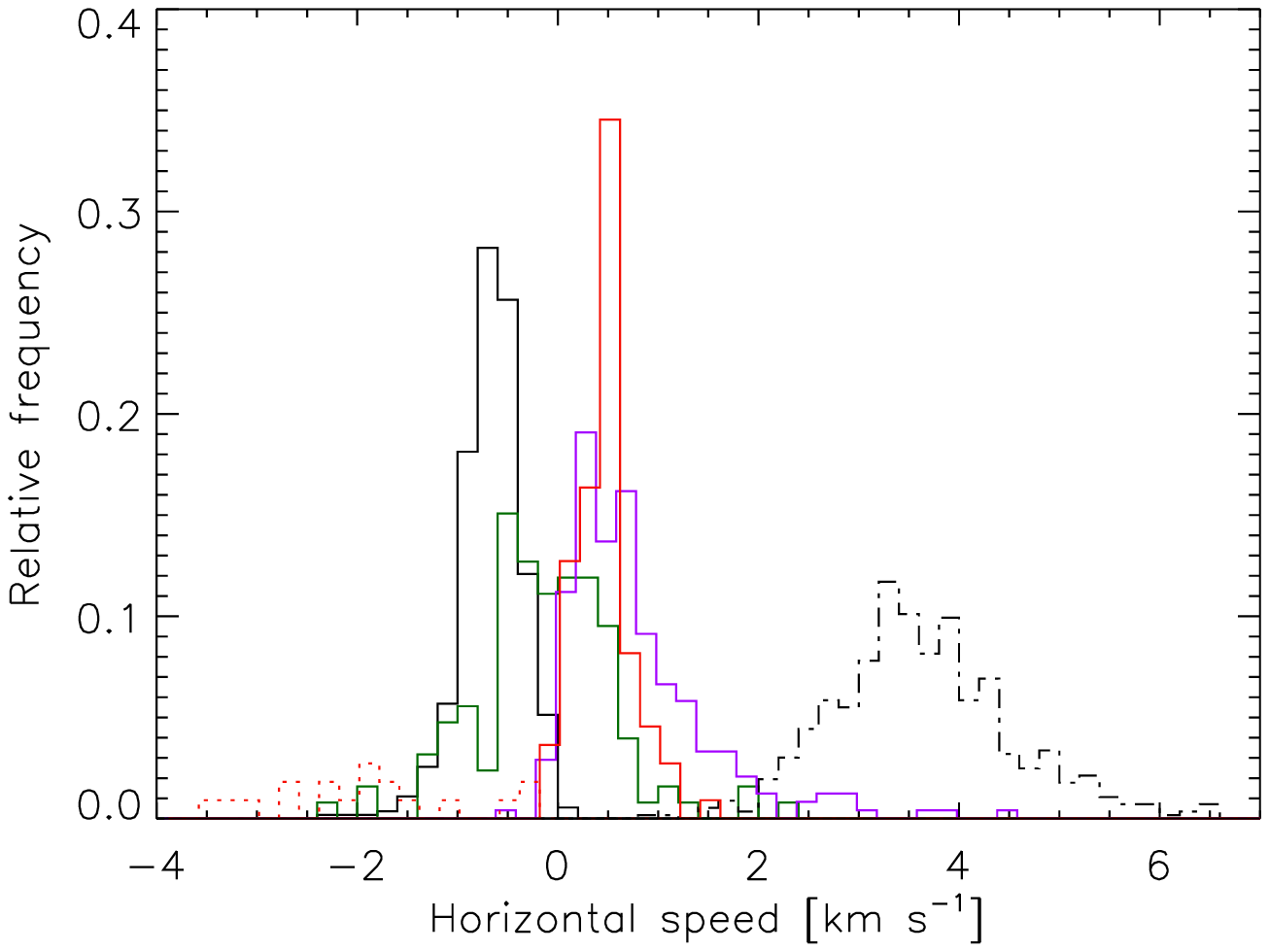}
\includegraphics[width=0.5\textwidth]{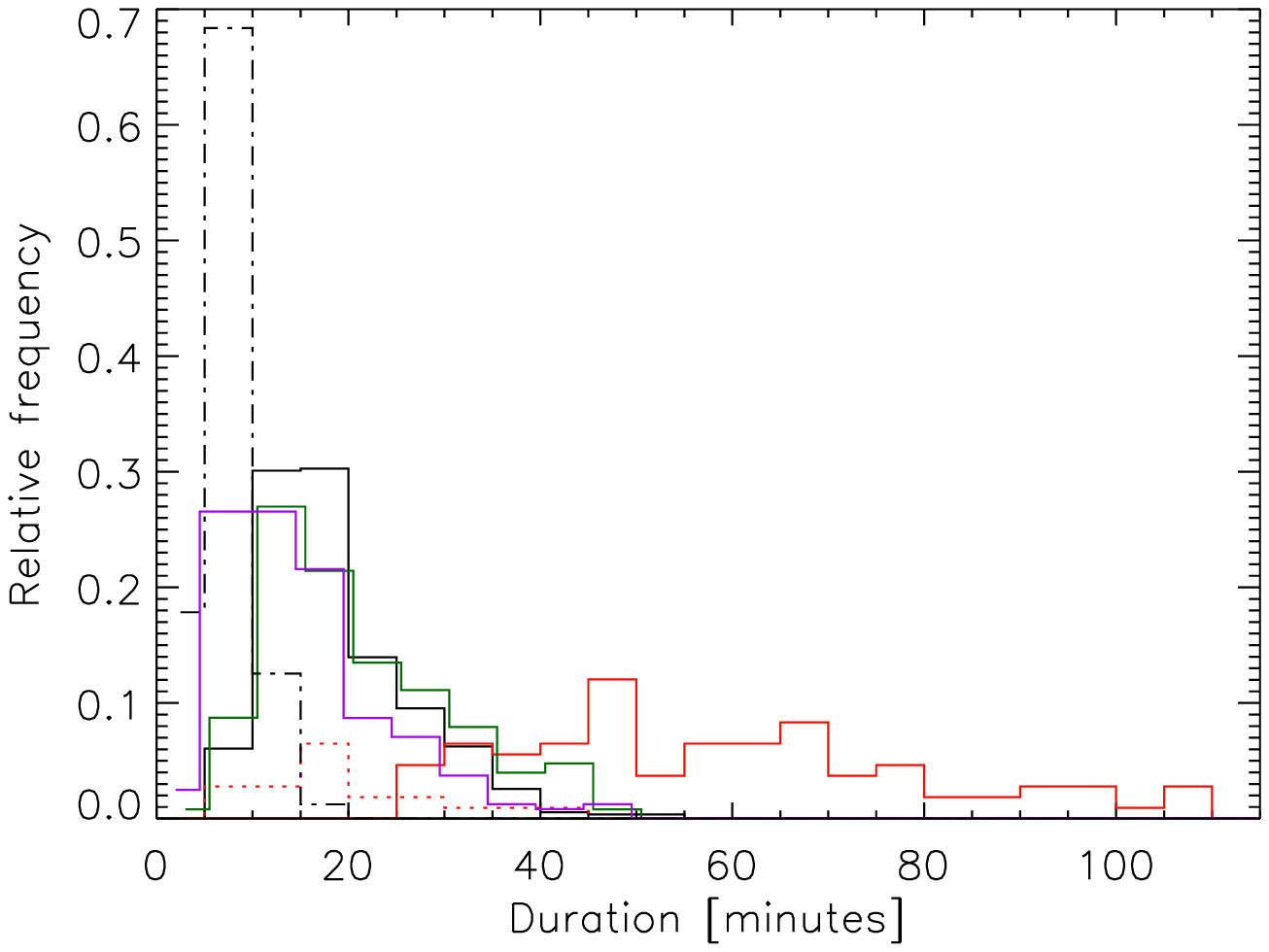}
\caption{Histograms of horizontal speeds ({\it top}) and durations (lifetimes) of the features ({\it bottom}). Black lines represent PGs in the inner half of the penumbra (solid line) and structures in dark bodies of filaments (dash-dot), green lines represent PGs in the outer half of the penumbra, red lines represent the motions of the outer penumbral border, both expansion and stationary (solid) and contraction (dotted), and purple  lines represent GBPs.
\label{fig:histo}}
\end{figure}

Histograms of horizontal speeds and durations are displayed in the top and bottom panels of Fig.~\ref{fig:histo}, respectively. Speeds of inward-moving PGs in the inner half of the penumbra show a narrow peak around $-0.7$~km\,s$^{-1}$ (solid black line). In the outer half of the penumbra, PGs move inwards and outwards with speeds of approximately $-1$ to 1~km\,s$^{-1}$ (green line), and some of them are stationary. The broad peak (dash-dot black) with high positive speeds corresponds to structures in dark filaments' bodies. The histogram of displacements of the outer penumbral border (red) has a strong peak at 0.5~km\,s$^{-1}$, corresponding to expansion, and some scattered contraction speeds (red dots) in the range $-3.5$ to $-0.5$~km\,s$^{-1}$. The speeds of GBPs are also mostly concentrated around 0.5~km\,s$^{-1}$, but their histogram (purple) has a tail of higher speeds of up to 3--4~km\,s$^{-1}$.
The histograms of the duration of PGs (black and green) and GBPs (purple) have similar asymmetric shapes, with maxima at shorter durations of 15 and 10~minutes, respectively. The time periods of motions of the outer penumbral border have a very broad range, but the contractions (red dots) clearly take much less time than the expansion and stationary periods (red lines).

The speeds of horizontal motions can be expected to depend on the positions of features in the penumbra. We defined a scale of relative distances from the umbra--penumbra border, setting zero at the inner and one at the outer penumbral boundaries. We defined the positions of the two boundaries by means of intensities averaged over the whole time period of 365 minutes for each individual time slice, adopting the values of 0.35\,$I_{\rm phot}$ for the inner and 0.90\,$I_{\rm phot}$ for the outer penumbral borders. The relative distance of each feature was then calculated from an average of the starting and ending horizontal coordinates of the line drawn over the feature's trace in the time slice.

\begin{figure}[t]\centering
\includegraphics[width=0.5\textwidth]{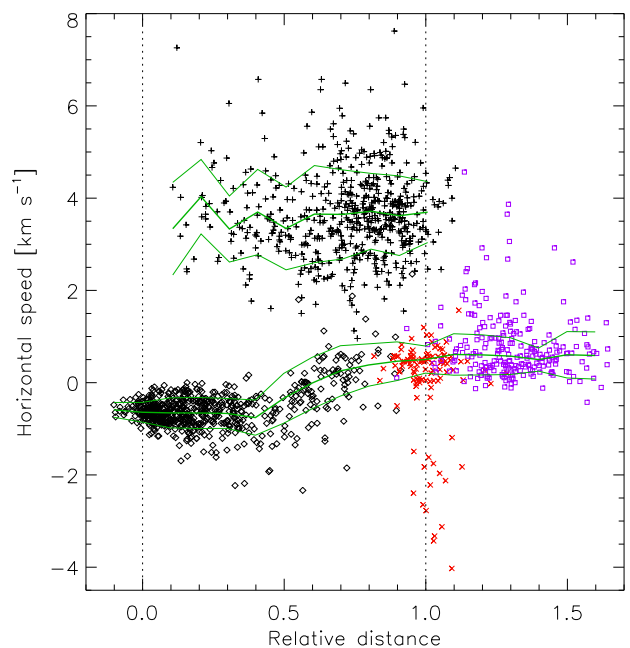}
\caption{Scatter plot of horizontal speeds versus relative distances of features from the umbra--penumbra border. Penumbral grains are marked by diamonds, structures in dark bodies of filaments by plus signs, motions of the outer penumbral border by red multiplication signs, and GBPs by purple squares. 
Green lines show average speeds together with standard deviations. Distances 0 and 1 correspond to time-averaged positions of the inner and outer penumbral borders, respectively.
\label{fig:scatplot}}
\end{figure}

A scatter plot of horizontal speeds versus relative distances from the umbra--penumbra border is displayed in Fig.~\ref{fig:scatplot}. We show the mean values and standard deviations of speeds calculated in 0.1 wide sections of the relative distance, $d$ (green lines).
The majority of measured PGs (black diamonds) are located in the inner third of the penumbra. These PGs move inwards, and their speeds do not depend on $d$. The speeds of PGs observed in the outer penumbra have, despite a large scatter, a clear trend of increasing with increasing $d$. The mean speed increases from $-0.7$~km\,s$^{-1}$ ($d \leq 0.4$) to 0.4~km\,s$^{-1}$ at $d = 0.8$. The dominant direction of PG motions changes from inwards to outwards at $d \simeq 0.6$. The mean outward speed of 0.5~km\,s$^{-1}$ is typical for the expansion periods of the outer penumbral border (red crosses). The major part of GBPs (purple squares) move away from the spot with a mean speed of 0.6~km\,s$^{-1}$; outlying speeds higher than 1.6~km\,s$^{-1}$, caused by local granular motions, are not included in this average. Moving structures in the dark bodies of filaments form a completely different system of largely scattered speeds. They are detected in the whole penumbra, and the majority of them appear at $d > 0.6$. Their mean speed, $3.6 \pm 0.9$~km\,s$^{-1}$, is independent of the relative position in the penumbra. 

\section{Discussion and conclusions} \label{sec:discuss}

We have measured horizontal motions in the penumbra of a large stable sunspot. Speeds and duration times of PGs, dark bodies of penumbral filaments, GBPs near the sunspot, and the repeated expansion and contraction of the outer penumbral border were measured visually in time slices created from red-continuum and G-band intensity movies of high spatial (90~km) and temporal (20.1~s) resolutions.

The presented time-slice method has a limitation because the measured features can move from one slice to another. It can be seen in the movies that when the penumbral filaments are curved, some features, particularly PGs in the inner penumbra, can cross two to four slices. This obviously results in an underestimate of the lifetimes but also of the speeds because the features do not move parallel to the slices. Nonetheless, the majority of filaments appear radially elongated with minimal deviations, so the statistical analysis of many measurements can provide realistic results. Also, the visual measurement is subjective to some extent because of the visual selection of measured traces of features and the manual drawing of lines over them. On the other hand, the time-slice method and visual recognition has the advantage of avoiding numerous false and noisy identifications of features, which often appear in the results of automated feature tracking.

The apparent horizontal-motion speeds and durations of PGs found in this work are in good agreement with those obtained previously from feature tracking \citep{Zhang2013, SBSIII1999, SSIV2001}, including the change of prevailing direction of motions from inwards to outwards at the relative distance $d \simeq 0.6$. The mean speed increases gradually from negative to positive values between $d = 0.4$ and 0.8 and matches the mean expansion speed of the outer penumbral border well. Likewise, the mean outflow speed of magnetic elements represented by GBPs in the sunspot moat is similar to the mean expansion speed of the penumbral border (Fig.~\ref{fig:scatplot}).

Speeds detected in the dark bodies of penumbral filaments are consistent with previous measurements of proper motions of Evershed clouds \citep{Shineetal1994, Cabreraetal2007}. We have to note that these speeds may be underestimated in our work because the subsonic filter used to suppress five-minute oscillations and residual seeing effects has a cutoff at 4~km\,s$^{-1}$, causing a reduction in observed speeds. The outward motions in the bodies of filaments can be expected to be associated with the Evershed flow.

According to \citet{Tiwari2013}, penumbral filaments represent magnetoconvective cells distributed everywhere in the penumbra and surrounded by a magnetic field of spines, the inclination of which gradually increases with increasing distance from the umbra. In the inner penumbra, the magnetic field in the filament's head (PG) is more inclined to the normal than that in its surroundings, but the opposite is true in the middle and outer penumbrae, as shown in Fig.~6 of \citet{Tiwari2013}. The same figure also shows that the magnetic field in a PG is weaker than the surrounding one in the inner penumbra but that both fields are comparable in the middle and outer penumbrae.

We suggest the following tentative scenario, which, however, needs to be proven by numerical simulations that are beyond the scope of this paper. Rising hot plasma surrounded by a stronger and less inclined magnetic field may adapt its trajectory to be more vertical, which would lead to the inward motion of PGs in the inner penumbra. Oppositely, in the outer penumbra the rising hot plasma in the filament's head is dragged by the surrounding, more horizontal magnetic field such that its crossing point with the visible surface (PG) moves outwards. The outer penumbral border is formed by the ends of filaments. Its repeated expansion with speeds comparable to that of outward-moving PGs can be tentatively explained if we assume that the filaments (magnetoconvective cells) in the outer penumbra move outwards as a whole. The periods of expansion are interrupted by the formation and growth of granules close to the penumbral border.

\begin{acknowledgements}

We thank the referee for constructive comments, which helped us to improve the paper.
We express our thanks to C. M\"ostl, R. Kever, and R. Henderson for assisting in the observations and J. Jur\v{c}\'{a}k for discussion. This work was done under the institutional support ASU:67985815 of the Czech Academy of Sciences. The Swedish 1-m Solar Telescope is operated on the island of La Palma by the Institute for Solar Physics of the Royal Swedish Academy of Sciences in the Spanish Observatorio del Roque de los Muchachos of the Instituto de Astrof\' {\i}sica de Canarias. 
\end{acknowledgements}


\bibliographystyle{aa}
\bibliography{bibliography1}

\end{document}